\documentstyle[preprint,aps,epsf,floats]{revtex}

\begin{document}
\tighten
\def\deut{d}
\def\rat{\aleph}
\def\rat{\delta}
\def\rat{\beta}
\def\si{{}^1\kern-.14em S_0}
\def\siii{{}^3\kern-.14em S_1}
\def\diii{{}^3\kern-.14em D_1}
\def\pone{{}^3\kern-.14em P_1}
\def\pzero{{}^3\kern-.14em P_0}
\def\ptwo{{}^3\kern-.14em P_2}
\newcommand{\gsim}{\raisebox{-0.7ex}{$\stackrel{\textstyle >}{\sim}$ }}
\newcommand{\lsim}{\raisebox{-0.7ex}{$\stackrel{\textstyle <}{\sim}$ }}
\def\pislash{ {\pi\hskip-0.6em /} }
\def\pislashsmall{ {\pi\hskip-0.375em /} }
\def\pslash{p\hskip-0.45em /}
\def\nopi{ {\rm EFT}(\pislash) }
\def\nopit{ {\rm dEFT}(\pislash) }
\def\nopith{ {\rm dEFT}(\pi_h) }
\def\Ltwo{ {^\pislashsmall \hskip -0.2em L_2 }}
\def\Lone{ {^\pislashsmall \hskip -0.2em L_1 }}
\def\CQuad{ {^\pislashsmall \hskip -0.2em C_{\cal Q} }}
\def\Czero{ {^\pislashsmall \hskip -0.2em C_{0}^{(\siii)} }}
\def\Czeromone{ {^\pislashsmall \hskip -0.2em C_{0,-1}^{(\siii)} }}
\def\Czerozero{ {^\pislashsmall \hskip -0.2em C_{0,0}^{(\siii)} }}
\def\Czeroone{ {^\pislashsmall \hskip -0.2em C_{0,1}^{(\siii)} }}
\def\Ctwo{ {^\pislashsmall \hskip -0.2em C_{2}^{(\siii)} }}
\def\Ctwomtwo{ {^\pislashsmall \hskip -0.2em C_{2,-2}^{(\siii)} }}
\def\Ctwomone{ {^\pislashsmall \hskip -0.2em C_{2,-1}^{(\siii)} }}
\def\Cfour{ {^\pislashsmall \hskip -0.2em C_{4}^{(\siii)} }}
\def\CSDzero{ {^\pislashsmall \hskip -0.2em C_0^{(sd)} }}
\def\CSDtwotwotwo{ {^\pislashsmall \hskip -0.2em C_{2,-2}^{(sd)} }}
\def\CSDzeromone{ {^\pislashsmall \hskip -0.2em C_{0,-1}^{(sd)} }}
\def\CSDzerozero{ {^\pislashsmall \hskip -0.2em C_{0,0}^{(sd)} }}
\def\CSDtwoone{ {^\pislashsmall \hskip -0.2em \tilde C_2^{(sd)} }}
\def\CSDtwotwo{ {^\pislashsmall \hskip -0.2em C_2^{(sd)} }}
\def\CSDzerotwo{ {^\pislashsmall \hskip -0.2em C_{0,0}^{(sd)} }}
\def\LX{ {^\pislashsmall \hskip -0.2em L_X }}
\def\CSDfour{ {^\pislashsmall \hskip -0.2em C_4^{(sd)} }}
\def\CSDfourt{ {^\pislashsmall \hskip -0.2em \tilde C_4^{(sd)} }}
\def\CSDfourtt{ {^\pislashsmall \hskip -0.2em {\tilde{\tilde C}}_4^{(sd)} }}
\def\etasd{\eta_{sd} }
\def\ZCzeromone{ 
{_z \hskip -0.4em {^\pislashsmall \hskip -0.2em C_{0,-1}^{(\siii)} }}}
\def\ZCzerozero{ 
{_z \hskip -0.4em {^\pislashsmall \hskip -0.2em C_{0,0}^{(\siii)} }}}
\def\ZCzeroone{ 
{_z \hskip -0.4em {^\pislashsmall \hskip -0.2em C_{0,1}^{(\siii)} }}}
\def\ZCtwomtwo{ 
{_z \hskip -0.4em {^\pislashsmall \hskip -0.2em C_{2,-2}^{(\siii)} }}}
\def\ZCtwomone{ 
{_z \hskip -0.4em {^\pislashsmall \hskip -0.2em C_{2,-1}^{(\siii)} }}}
\def\ZCfourmthree{ 
{_z \hskip -0.4em {^\pislashsmall \hskip -0.2em C_{4,-3}^{(\siii)} }}}
\def\rCzeromone{ 
{_\rho \hskip -0.4em {^\pislashsmall \hskip -0.2em C_{0,-1}^{(\siii)} }}}
\def\rCzerozero{ 
{_\rho \hskip -0.4em {^\pislashsmall \hskip -0.2em C_{0,0}^{(\siii)} }}}
\def\rCzeroone{ 
{_\rho \hskip -0.4em {^\pislashsmall \hskip -0.2em C_{0,1}^{(\siii)} }}}
\def\rCtwomtwo{ 
{_\rho \hskip -0.4em {^\pislashsmall \hskip -0.2em C_{2,-2}^{(\siii)} }}}
\def\rCtwomone{ 
{_\rho \hskip -0.4em {^\pislashsmall \hskip -0.2em C_{2,-1}^{(\siii)} }}}
\def\rCfourmthree{ 
{_\rho \hskip -0.4em {^\pislashsmall \hskip -0.2em C_{4,-3}^{(\siii)} }}}
\def\CPzero{ {^\pislashsmall \hskip -0.2em C^{(\pzero)}_2  }}
\def\CPone{ {^\pislashsmall \hskip -0.2em C^{(\pone)}_2  }}
\def\CPtwo{ {^\pislashsmall \hskip -0.2em C^{(\ptwo)}_2  }}

\def\Journal#1#2#3#4{{#1} {\bf #2}, #3 (#4)}

\def\NCA{\em Nuovo Cimento}
\def\NIM{\em Nucl. Instrum. Methods}
\def\NIMA{{\em Nucl. Instrum. Methods} A}
\def\NPB{{\em Nucl. Phys.} B}
\def\NPA{{\em Nucl. Phys.} A}
\def\NP{{\em Nucl. Phys.} }
\def\PLB{{\em Phys. Lett.} B}
\def\PRL{\em Phys. Rev. Lett.}
\def\PRD{{\em Phys. Rev.} D}
\def\PRC{{\em Phys. Rev.} C}
\def\PRA{{\em Phys. Rev.} A}
\def\PR{{\em Phys. Rev.} }
\def\ZPC{{\em Z. Phys.} C}
\def\SJP{{\em Sov. Phys. JETP}}
\def\SJNP{{\em Sov. Phys. Nucl. Phys.}}

\def\FBS{{\em Few Body Systems Suppl.}}
\def\IJMP{{\em Int. J. Mod. Phys.} A}
\def\UJP{{\em Ukr. J. of Phys.}}
\def\CJP{{\em Can. J. Phys.}}
\def\SCI{{\em Science} }
\def\AST{{\em Astrophys. Jour.} }
\def\tran{dibaryon}
\def\trans{dibaryons}
\def\Tran{Dibaryon}
\def\Trans{Dibaryons}
\def\TRANS{DIBARYONS}
\def\yt{y}

\preprint{\vbox{
\hbox{ NT@UW-02-010}
}}
\bigskip
\bigskip

\title{Pions in the Pionless Effective Field Theory}

\author{{\bf Silas R. Beane} and {\bf Martin J. Savage}}
\address{Department of Physics, University of Washington, \\
Seattle, WA 98195. }
\maketitle

\begin{abstract}
  We show that processes involving pions that remain very near their mass-shell
  can be reliably computed in the pionless effective field theory, with the
  pion integrated in as a heavy field. 
As an application, we compute the
$\pi$-deuteron scattering amplitude near threshold to next-to-leading order
  in the momentum expansion.  
This amplitude is formally dominated by an infrared 
logarithm of the 
form $\log\left(\gamma/m_\pi\right)$,
where $\gamma$ is the deuteron binding momentum, 
and $m_\pi$ is the mass of the  pion.
The coefficient of this logarithm is 
determined by the S-wave pion-nucleon scattering lengths.
\end{abstract}

\vskip 2in

\leftline{May 2002.}
\vfill\eject


\section{Introduction}

For many processes involving pions scattering from multi-nucleon systems, the
kinematics are such that the pions must be included as dynamical particles, as
they can carry energy and momentum that is not small compared to their rest
mass, and can be far off-shell.  However, in the case of low-energy
pion-deuteron ($\pi d$) scattering one has a situation where the pions can
carry energy and momentum that puts them on or near their mass-shell, both on
external legs and internally in some diagrams. 
Furthermore, given the small binding energy of
the deuteron, the two nucleons in the deuteron are off their mass-shell by an
amount that is far less than the pion mass. Therefore, we can construct an
effective field theory (EFT) where non-relativistic nucleons interact via
contact operators, and pions near their mass-shell are included as massless
particles that generate a coulomb-like potential.  The
construction of this low-energy EFT is similar to the development of the
heavy-quark effective theory (HQET)~\cite{Isgur:vq,Georgi:1990um} and non-relativistic QCD
(NRQCD)~\cite{Grinstein:1997gv,Luke:1997ys,Griesshammer:1998fh}.

The EFT relevant to momentum transfers much less than the pion mass, in which
nucleons interact via contact operators, is known as the pionless EFT, or
$\nopi$. The power-counting rules in $\nopi$ are now an established part of the
nuclear physics literature and can be found in several review
articles~\cite{Ioffe}, together with reviews of the substantial array of
phenomena that have been studied in $\nopi$.  A recently developed variant of
$\nopi$ is particularly suited to high-order, 
precision calculations~\cite{Beane:2000fi}. In this EFT, known as dibaryon
pionless EFT, or $\nopit$, the
$\siii$ NN effective range parameter, $r_3$, is taken to be of order
$1/Q$ in the power counting, where $Q$ is a small momentum~\cite{Ka99}.  This
naturally leads to the use of \tran{} fields to account for nonperturbative
enhancements~\cite{Ka97}.  In $\nopit$, operators are ordered according to
powers of the total center-of-mass energy~\cite{Beane:2000fi}.  Unlike $\nopi$,
the higher-dimension operators involving external fields are not renormalized
by the S-wave strong interactions, and therefore do not scale with inverse
powers of the renormalization scale. Thus, naive dimensional analysis of these
operators is sufficient to estimate their contribution to a given process.
Due to these technical advantages, we will work with $\nopit$ in this paper.

\section{Integrating in the Pions}

Given that both $\nopi$~\cite{Ch99,KSW,vanKolck:1998bw,Bedaque:1997qi} 
and $\nopit$~\cite{Beane:2000fi,Gabbiani:2001yh} 
have been studied in detail, the remaining ingredient is the pion.
In analogy with HQET and 
heavy-baryon chiral perturbation theory (HB$\chi$PT), 
for processes involving the pions
very near their mass-shell it is convenient to perform field redefinitions 
to remove the classical trajectory of a pion and use nonrelativistic
kinematics.
In the limit that the nucleon mass is infinitely larger than the pion mass,
and the deuteron binding energy is taken to be much smaller than any other
scale in the problem, the pion kinetic energy
in the deuteron rest frame, $T_\pi$,  
can be taken to be a constant even in loop diagrams.
Furthermore, the theory can be constructed in terms of 
the annihilation 
and creation operators for the pions.
In loop diagrams the typical pion residual kinetic energy is of order
$\gamma^2/ M_N$, while its momentum is of order $\gamma$, 
where $\gamma$ is the deuteron binding momentum.
Thus, as is the case in NRQCD, it is possible and necessary to treat 
the residual energy operator in
perturbation theory. 
Working in the isospin limit, the leading-order (LO) 
Lagrange density describing non-interacting
``heavy pions'', $\pi_h$, 
is~\footnote{We could have chosen a different normalization for 
$\pi_h$ such that $\tilde \pi_h = \sqrt{2 m_\pi}\pi_h$,
in which case the kinetic term would be 
\begin{eqnarray}
{\cal L } & = & 
\pi^{+\dagger}_h \left(\  T_\pi + {\nabla^2\over 2 m_\pi}\ \right) \pi^+
\ +\ ...
\ \ \ , \nonumber
\end{eqnarray}
which is a more familiar form for the non-relativistic Lagrange density.
The ellipses denote the kinetic-energy terms for the $\pi^-$ and $\pi^0$.
}
\begin{eqnarray}
{\cal L } & = & 
{\rm tr}\left[\ M_h^\dagger\ \left( 2 m_\pi T_\pi + \nabla^2\ \right) M_h\
\right]
\ \ \ ,
\label{eq:lagpiHQ}
\end{eqnarray}
where $T_\pi$ is the kinetic energy of the pion
and where $m_\pi$ is the pion mass.
The subscript ``$h$'' denotes  a heavy field, and 
we have defined the meson matrices
\begin{eqnarray}
M_h & = & \left(\matrix{ 
\pi^0_h/\sqrt{2} & \pi^+_h\cr\pi^-_h & -\pi^0_h/\sqrt{2}}
\right)
\qquad , \qquad
M_h^\dagger \ =\ \left(\matrix{ 
\pi^{0\dagger}_h/\sqrt{2} & \pi^{-\dagger}_h\cr \pi^{+\dagger}_h & 
-\pi^{0\dagger}_h/\sqrt{2}}
\right)
\ \ \ \ .
\label{eq:HQmesons}
\end{eqnarray}
There is an implied summation over all possible pions with
kinetic energy $T_\pi$, just as there is a summation over all possible
four-velocities in HQET.
The operator $\pi_h$ annihilates a pion while the operator 
$\pi_h^\dagger$ creates a 
pion.
The propagator for the pion has the form
\begin{eqnarray}
D_{T_\pi}(q^0, {\bf q}) & = & {i\over 2 m_\pi T_\pi - |{\bf q}|^2 + i \epsilon}
\ \ \ ,
\label{eq:piprop}
\end{eqnarray}
which is independent of $q^0$, and reduces down to that of a 
``potential photon'' in NRQED or a ``potential gluon'' in NRQCD, 
(i.e. $-i/(|{\bf q}|^2 -  i\epsilon)$) in the limit that the 
pion has vanishing kinetic energy, $T_\pi\rightarrow 0$.

We describe deuteron properties using $\nopit$, in which the scattering length
{\it and} effective range in the $\siii$-channel are assumed to be 
unnaturally large, and to
scale like $1/Q$ in the power-counting~\cite{Beane:2000fi}. 
The Lagrange density describing the dynamics of the nucleons and 
\tran{}, $t^j$, in the
$\siii$ channel is
\begin{eqnarray}
{\cal L}_t & = & 
N^\dagger \left[ i\partial_0 + {\nabla^2\over 2 M_N}\right] N
\ -\ 
t^{\dagger}_j \left[  i\partial_0 + {\nabla^2\over 4 M_N} - \Delta \right] t^j
\ -\  \yt \left[\  t^\dagger_j \ N^T P^j N\ +\ {\rm h.c.} \right]
\ \ \ ,
\label{eq:trandef}
\end{eqnarray}
where $y$ is the coupling between nucleons in the $\siii$ channel and the 
$\siii$-\tran.
The spin-isospin projector for the $\siii$ channel is
\begin{eqnarray}
P^i & \equiv &  {1\over \sqrt{8}} \sigma_2\sigma^i\  \tau_2
\ \ \ , 
\ \ \  {\rm Tr} \left[ P^{i\dagger} P^j \right]\  =\ {1\over 2} \delta^{ij}
\ \ \ .
\label{eq:progT}
\end{eqnarray}
%
\begin{figure}[!ht]
\vskip 0.15in
\centerline{{\epsfxsize=6.3in \epsfbox{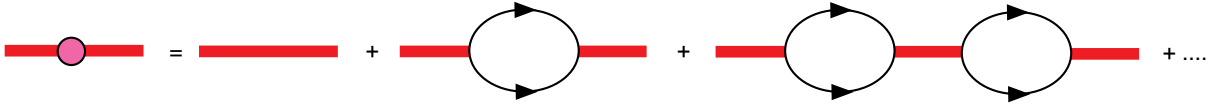}} }
\vskip 0.15in
\noindent
\caption{\it The dressed \tran{} propagator. 
The bare \tran{} propagator
is dressed by nucleon bubbles to all orders. 
Each diagram
counts as $Q^{-2}$ in the power-counting scheme.
}
\label{fig:trans}
\vskip .2in
\end{figure}
It is easy to show that this Lagrange density alone reproduces 
the NN scattering amplitude in the $\siii$-channel, with scattering length
$a_3$ and effective range $r_3$, when
\begin{eqnarray}
\yt^2 & = & {8\pi\over M_N^2 r_3}
\ \ ,\ \ 
\Delta\ =\ {2\over M_N  r_3} \left({1\over a_3} - \mu\right)
\ \ \ ,
\label{eq:coupfix}
\end{eqnarray}
where $\mu$ is the renormalization scale. This identification is made after
dressing the \tran{} propagator, as shown in Fig.~\ref{fig:trans}.
Higher order contributions to the scattering amplitude 
arising from the shape parameter
and higher order terms in the effective range expansion are understood to be included
perturbatively.

\begin{figure}[!ht]
\vskip 0.15in
\centerline{{\epsfxsize=4.3in \epsfbox{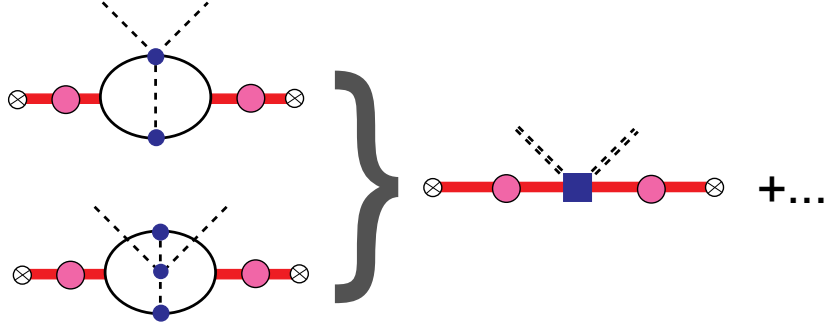}} }
\vskip 0.15in
\noindent
\caption{\it Diagrams with interaction vertices that do not conserve pion number
(left) appear as contact interactions (right) in the heavy-pion EFT. 
The single (double) dashed lines represent ordinary (heavy) pions 
and the solid black lines represent nucleons. 
The solid bar with a blob is a 
dressed \tran{} field.}
\label{fig:heavypid3}
\vskip .2in
\end{figure}
In order to have the incoming and outgoing nucleons near their mass-shell and
also to have all the pions in a process near their mass-shell, there must
be an equal number of pions entering and leaving each interaction vertex.
As illustrated in Fig.~\ref{fig:heavypid3}, processes with a different number
of pions entering than leaving~\cite{Jerry,wein1,Beane:1998y}  will be
represented by local operators (on the scale of $m_\pi$) in the EFT.
One can assign a charge to the pions --not carried by the
nucleons-- which must be conserved in all processes in the EFT.
Interactions between a heavy pion and a nucleon are
described by a Lagrange density of the form
\begin{eqnarray}
{\cal L } & = & 
c_0 (T_\pi)\ N^\dagger N \ {\rm tr}\left[ M_h^\dagger M_h \right]
\ +\ 
c_1 (T_\pi) \  N^\dagger \tau^a N\  
{\rm tr}\left[ \tau^a \left[M_h^\dagger , M_h\right]\  \right] \ \ .
\label{eq:Npi}
\end{eqnarray}
As there are no loop contributions from the heavy pion 
fields~\footnote{For the bubble contributions to the
$\pi_h N$ scattering amplitude,
one can close the contour in the upper half of the complex energy plane to
avoid enclosing a pole in the energy integration.
Thus all bubble contributions vanish. This generalizes to all bubble diagrams
involving heavy pions.}, 
the coefficients $c_{0,1} (T_\pi) $ represent the complete
pion-nucleon ($\pi N$) scattering amplitude evaluated at the center of mass
energy $T_\pi$.
Hence, these coefficients are related to the parameters in the effective range
expansion for $\pi N$ scattering via
\begin{eqnarray}
c_0 (T_\pi) & = &\  
-4\pi \left(1+\rat\right) \ {1\over k\cot\delta_{\pi N}^{+} - i
  \sqrt{2 m_\pi T_\pi}}
\ \ ,\ \ \nonumber\\  
c_1 (T_\pi)\ &=& 
\  2\pi \left(1+\rat\right) \  {1\over k\cot\delta_{\pi N}^{-} - i
  \sqrt{2 m_\pi T_\pi}}
\ \ \ \ ,
\end{eqnarray}
where the effective range expansion for $k\cot\delta_{\pi N}^{\pm}$ has the
form
\begin{eqnarray}
k\cot\delta_{\pi N}^{\pm} & = & -{1\over a^\pm} \ +\ ...
\ \ \ ,
\end{eqnarray}
where $a^\pm$ are the scattering lengths 
in the s-wave isoscalar, $(+)$, and isovector channels,$(-)$,
and $\delta_{\pi N}$ are the phase shifts for $\pi N$ scattering.
The ellipses denote the contributions from the effective range terms and
higher.
We have retained the quantity $\rat\equiv{m_\pi/M_N}$ in these expressions as
it arises from a trivial kinematic relation between the scattering amplitude
and the Lagrange density when $T_\pi=0$. 
However, it is important to keep in mind that the there will be contributions
from higher orders in the expansion that have factors of $\beta$ 
associated with them due to the explicit $1/M_N$ expansion.
At threshold, these coefficients reduce down to 
\begin{eqnarray}
c_0 (0) & = &\  4\pi \left(1+\rat\right) \ a^+
\ \ ,\ \ 
c_1 (0)\ = \  -2\pi \left(1+\rat\right) \ a^- 
\ \ \ \ .
\end{eqnarray}

To determine the coupling between the heavy pions and the \tran{} field
in the $\siii$-channel, we
realize that in the limit that the pions only interact once with the nucleons,
and using the fact that the deuteron is isoscalar, the only contribution to 
the $\pi d$ amplitude is from the operator with coefficient $c_0$.
Furthermore, this contribution must be absolutely normalized, as the pions 
couple to the baryon-number operator $N^\dagger N$.
This gives rise to a uniquely determined interaction between the heavy pions
and the \tran{} field, described by
\begin{eqnarray}
{\cal L} & = & 
\ -\ 2 c_0 (T_\pi)\ 
t_j^\dagger\ t^j\ {\rm tr}\left[ M_h^\dagger M_h \right] 
\ \ \ \ .
\label{eq:lotrani}
\end{eqnarray}
As the interactions induced by the heavy pions are subleading in the momentum
expansion, it is anticipated that the modifications to 
eq.~(\ref{eq:lotrani}) occur only at  subleading order.
Indeed, at NLO in the EFT one expects a contribution from a
counterterm whose coefficient is not dictated by baryon-number conservation.
We write this contribution as
\begin{eqnarray}
{\cal L} & = & {\eta (T_\pi, \mu)\over \Lambda M_N r_3}\ t_j^\dagger\ t^j\ 
 {\rm tr}\left[ M_h^\dagger M_h \right]
\ \ \ ,
\label{eq:nlotrani}
\end{eqnarray}
which scales like $\sim Q$ in the $\nopit$ power-counting. 
Here $\Lambda\sim m_\pi$ and $\mu$ is the renormalization scale.
In general, the coefficient function 
$\eta (T_\pi ,\mu)$ will have both real and imaginary
parts, as bubble diagrams involving the pion vanish.
The factors of 
$M_N$, $\Lambda$ and $r_3$ enter as per the power-counting appropriate for
\tran{} fields~\cite{Beane:2000fi}.

\section{$\pi \deut$ Scattering}

\begin{figure}[!ht]
\centerline{{\epsfxsize=4in \epsfbox{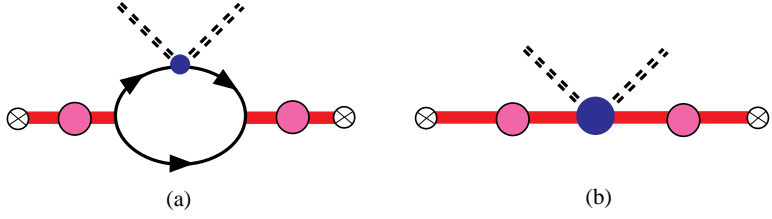}} }
\vskip 0.15in
\noindent
\caption{\it Graphs contributing to $\pi d$ scattering 
at LO. The dark ovals in diagrams (a) and (b)
are interaction vertices taken 
from eq.~(\ref{eq:Npi}) and
from eq.~(\ref{eq:lotrani}), respectively.}
\label{fig:heavypid2}
\vskip .2in
\end{figure}

At low-momentum, the LO amplitude for  $\pi d$ scattering 
receives contributions from the diagrams shown in 
Fig.~\ref{fig:heavypid2}.
There is the one-loop diagram with a
single insertion of the $\pi N$ scattering amplitudes
in eq.~(\ref{eq:Npi}),
and the local $\pi d$ operator given in  
eq.~(\ref{eq:lotrani}).
%
At LO, the operator that is inserted in the deuteron is a charge
operator, and consequently it is absolutely normalized at zero-momentum
transfer, and the form factor is identical to that of the deuteron charge form
factor~\cite{Beane:2000fi} at LO. 
Therefore, at LO, the $\pi d$ scattering amplitude is 
\begin{eqnarray}
{\cal A}^{(LO)}_{\pi d}(T_\pi, |{\bf k}|)    & = & 
2\  c_0 (T_\pi)\ {\gamma r_3\over1-\gamma r_3}
\ \left[\ {4\over r_3 |{\bf k}|} \tan^{-1}\left({ |{\bf k}|\over
      4\gamma}\right)\ -1\ \right]
\ \ \ ,
\label{eq:LOmom}
\end{eqnarray}
where $\gamma = \sqrt{M_N B}$ is the deuteron binding momentum, 
with $B$ the deuteron binding energy.
The momentum transfer to the deuteron is ${\bf k}$.
As expected, this form factor reduces to 
${\cal A}^{(LO)}_{\pi d} = 2\ c_0 (T_\pi)$, 
proportional to the isoscalar charge of the deuteron,
in the limit ${\bf k}\rightarrow 0$.

\begin{figure}[!ht]
\centerline{{\epsfxsize=4in \epsfbox{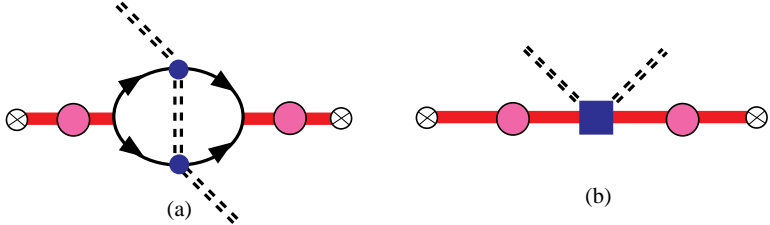}} }
\vskip 0.15in
\noindent
\caption{\it 
Graphs contributing to $\pi$-d scattering at NLO. 
The logarithmically  
divergent two-loop graph arising from heavy pion exchange, diagram  (a), 
is renormalized by the  local $\pi d$ operator 
given in eq.~\ref{eq:nlotrani},
denoted by the solid square in diagram  (b).}
\label{fig:heavypid}
\vskip .2in
\end{figure}

At next-to-leading-order (NLO) in the EFT expansion, there is a contribution
from a two-loop diagram (rescattering) involving a heavy-pion propagating
between the nucleons, and from a local $\pi d$ operator, 
as shown in Fig.~\ref{fig:heavypid}.
The amplitude at NLO is
\begin{eqnarray}
& & {\cal A}^{(NLO)}_{\pi d}(T_\pi, |{\bf k}|) \ =\ 
{\gamma\over (1-\gamma r_3)}
\left[\  \tilde\eta (T_\pi, \mu) 
\right. \nonumber\\ & & \left.
\ -\  {1\over 2\pi} 
\left[c_0 (T_\pi)^2-8 c_1 (T_\pi)^2\right]\ 
\left(\ 
\log\left({4\left(2\gamma - i \sqrt{ 2 m_\pi T_\pi}\right)^2  
+ |{\bf k}|^2\over 4\mu^2}\right) \right.\right. \nonumber\\ & & \left.\left. 
\qquad\qquad\qquad\qquad
\ +\ {4{\left(2\gamma - i \sqrt{ 2 m_\pi T_\pi}\right)}\over |{\bf k}|} 
\tan^{-1}\left({|{\bf k}|\over {2\left(2\gamma - i \sqrt{ 2 m_\pi T_\pi}
\right)}}\right)
\ -\ 2\ 
\right)
\ \right]
\ ,
\label{eq:twomom}
\end{eqnarray}
where we have used the integrals given in the Appendix of Ref.~\cite{SS01}.
The two-loop diagram is logarithmically
divergent and is regulated in the IR by the deuteron binding energy
for incident pions with vanishing kinetic energy.
The $\mu$-dependence of the counterterm, $\tilde\eta (T_\pi, \mu)$, 
must exactly
compensate the $\mu$-dependence of the loop diagram.
We have used the relation $\tilde\eta (T_\pi, \mu)
= \eta (T_\pi, \mu) /(\Lambda M_N)$.
Given that the scale where $\nopit$ 
is expected to breakdown is $\Lambda\sim m_\pi$,
we choose to renormalize the theory at $\mu\sim m_\pi$.

In the $T_\pi, |{\bf k}|\rightarrow 0$ limit this amplitude reduces to 
\begin{eqnarray}
{\cal A}^{(NLO)}_{\pi d} & = & 
{\gamma\over (1-\gamma r_3)}
\left[\  \tilde\eta (0, \mu)\ -\  {1\over\pi} 
\left[\ c_0 (0)^2-8 c_1 (0)^2\ \right]\ 
\log\left({2\gamma\over\mu}\right)\ \right]
\ \ \ ,
\label{eq:twozero}
\end{eqnarray}
It is important to notice that there is a contribution that 
is non-analytic in the deuteron binding momentum,
that is logarithmically enhanced,
$\sim \log\left(\gamma/ m_\pi\right)$, 
arising from the two-loop diagram.
This logarithm was found previously in the work of 
Borasoy and Grie\ss hammer~\cite{Borasoy:2001gq}.

The scale dependence of the logarithm determines the naive-dimensional
analysis (NDA) estimate for $\tilde\eta (0,m_\pi)$:
\begin{eqnarray}
| \tilde\eta (0, m_\pi) | & \sim & 
{1\over\pi}\  | c_0(0)^2-8 c_1(0)^2 |
\ \ \ .
\label{eq:NDA}
\end{eqnarray}
Note that the logarithmic divergence in the two-loop diagram 
modifies the NDA  analysis for the \tran{} operators of 
Ref.~\cite{Beane:2000fi}.
It is straightforward to show that other contributions to the
amplitude, both momentum-dependent and momentum-independent, 
are higher order in the
power-counting.  Therefore the sum of amplitudes given in 
eq.~(\ref{eq:LOmom}) and eq.~(\ref{eq:twomom}) constitute the complete 
amplitude up to NLO.

While we have constructed the EFT 
appropriate for describing the very-low energy behavior of $\pi d$ scattering,
there is additional information that we have at our disposal from
HB$\chi$PT. 
We know the chiral expansions of the coefficients $c_0$ and $c_1$;
at LO, $c_1(0)\sim m_\pi$ while $c_0 (0)\sim m_\pi^2$~\cite{ulf}.
Armed with this information we see that, in fact,
${\cal A}^{(LO)}_{\pi d}\sim c_0 (0)$ and 
${\cal A}^{(NLO)}_{\pi d}\sim c_1(0)^2$ are of the same order.  This does not
represent a breakdown of the low-energy EFT, but results from the LO amplitude
being anomalously small. Thus, given the anomalous behavior of the LO
amplitude, the leading contribution to low-energy $\pi d$ scattering is
\begin{eqnarray}
{\cal A}_{\pi d} & = & {\cal A}^{(LO)}_{\pi d} \ +\ {\cal A}^{(NLO)}_{\pi d}
\ +\ ...
\ \ \ ,
\label{eq:totamp}
\end{eqnarray}
as given in eq.~(\ref{eq:LOmom}) and eq.~(\ref{eq:twomom}).
The ellipses denote terms that are higher order in the momentum expansion,
suppressed by additional factors of $\gamma/m_\pi$.

To NLO, the $\pi d$ scattering length is given by
\begin{eqnarray}
{a}_{\pi d}={2\over{(1+\rat /2)}}&&\left[\ (1+\rat )\ a^+  \right. \nonumber\\
&& \left. -{{2\gamma}\over{1-\gamma r_3}}\ \left[\ 
(1+\rat )^2 \ 
\left[(a^{+})^2-2(a^{-})^2\right] \ 
\log\left({2\gamma\over{m_\pi}}\right)\ 
-{{\tilde\eta (0, {m_\pi})}\over{16\pi}}\ \right]\ 
\right] .
\label{eq:practpidscattlength}
\end{eqnarray}
It is interesting to compare this formula with that given in Ref.~\cite{eric}.
There, instead of the logarithm, one has the overlap of the coulomb-like pion
propagator between deuteron wave functions. Of course, since logarithms are
universal, the long-distance tail of any sensible deuteron wave function will
generate the infrared logarithm. The choice of different deuteron wave
functions corresponds to a different choice for the counterterm, 
${\tilde\eta (0, {m_\pi})}$. In principle, one may be able to determine
${\tilde\eta (0, {m_\pi})}$ uniquely using wave functions~\cite{beanewhat} 
computed in a fully consistent 
pionful EFT~\cite{weinbasic,Ordonez:1995rz,Epelbaum:1999dj,Beane:2001bc}. 

We can estimate ${a}_{\pi d}$ by taking ${\tilde\eta (0, {m_\pi})}=0$, which
corresponds to choosing the scale of the logarithm to be set by $m_\pi$. 
We use the Neuchatel-PSI-ETHZ (NPE) pionic hydrogen 
measurement~\cite{Schroder:uq}~\footnote{We do not use the
determinations of the scattering lengths in Ref.~\cite{Schroder:uq}
which include a measurement of 
${\rm Re}\left(a_{\pi d}\right)$~\cite{Hauser:1998yd}.
}
of the isovector $\pi N$ scattering length, 
$a^{-}_{\scriptstyle exp}=\left(-0.0905 \pm 0.0042\right)~m_\pi^{-1}$.
The isoscalar $\pi N$ scattering length, 
$a^{+}_{\scriptstyle exp}=\left(-0.0022 \pm 0.0043\right)~m_\pi^{-1}$
is consistent with zero~\cite{Schroder:uq} and therefore we will neglect it
in our estimate of ${a}_{\pi d}$.
Using $m_\pi = m_{\pi^+} = 139.57~{\rm MeV}$ and 
$M_N=(M_n + M_p)/2 = 938.92~{\rm MeV}$,
we find ${a}_{\pi d}= -0.019~m_\pi^{-1}$ as compared to the
NPE pionic deuterium measurement~\cite{Hauser:1998yd} of 
${\rm Re}\left({a}_{\pi d}^{\scriptstyle exp}\right)
=\left(-0.0261 \pm 0.0005\right)~m_\pi^{-1}$.
Therefore, we see that the leading non-analytic piece
provides the dominant contribution to $\pi d$ scattering.
However, as expected,
the contribution from the counterterm $\tilde\eta$ is an important 
contribution that cannot be discarded.
It is worth pointing out that an imaginary part of
${a}_{\pi d}^{\scriptstyle exp}$ will be induced at higher orders 
in the expansion.  This is consistent with the smallness of the measured value
of ${\rm Im}\left({a}_{\pi d}^{\scriptstyle exp}\right)
=\left(-0.0063 \pm 0.0007\right)~m_\pi^{-1}$~\cite{Hauser:1998yd}.

The $\pi d$ scattering amplitude can be expressed entirely 
in terms of the $\pi N$ and $\pi d$ scattering lengths,
and the difference $\tilde\eta (T_\pi, m_\pi) - \tilde\eta (0, m_\pi)$.
At the order to which we are working, this latter difference will be 
completely imaginary, and will be determined from matching conditions. 
In particular, the scattering amplitude must be unitary (we are not allowing
for energy transfer from the pion) and thus the optical theorem will
determine $\tilde\eta (T_\pi, m_\pi) - \tilde\eta (0, m_\pi)$, since 
$\tilde\eta$ is independent of the three-momentum transfer.
Moreover, this difference will be independent of the renormalization scale at this order.
Expanding out the scattering amplitudes in 
$\sim \sqrt{2 m_\pi T_\pi}/m_\pi$, we find 
\begin{eqnarray}
{\cal A}_{\pi d}(T_\pi, |{\bf k}|)&= &  
4\pi (1+\rat /2)\ a_{\pi d} 
\ +\ 
{{8\pi(1+\rat)}\over 1-\gamma r_3}
a^+ \left[\ {{4\gamma}\over{|{\bf k}|}} \tan^{-1}\left({ |{\bf k}|\over
      4\gamma}\right)\ -1 \ \right]
\nonumber\\
& & -i { 8\pi  (a^+)^2 \sqrt{2 m_\pi T_\pi} (1+\rat) \gamma r_3\over
1-\gamma r_3} 
 \left[\ {4\over r_3 {|{\bf k}|}} \tan^{-1}\left({ |{\bf k}|\over
      4\gamma}\right)\ -1 \ \right]
\nonumber\\
\ & +&\ 
{{8\pi\gamma (1+\rat)^2}\left[(a^{+})^2-2(a^{-})^2\right]\over 1-\gamma r_3}
\left[\ 
\log\left(
{{16\gamma^2}\over{4\left(2\gamma - i \sqrt{ 2 m_\pi T_\pi}\right)^2 + |{\bf k}|^2}}\right) 
\ \right. \nonumber\\ & & \left.
\ -\ {4{\left(2\gamma - i \sqrt{ 2 m_\pi T_\pi}\right)}\over |{\bf k}|} 
\tan^{-1}\left({|{\bf k}|\over {2\left(2\gamma - i \sqrt{ 2 m_\pi T_\pi}\right)}}\right)
\ +\ 2\ 
\right] 
\nonumber\\
& & + {\gamma\over 1-\gamma r_3}\left[\ \tilde\eta (T_\pi, m_\pi) - 
\tilde\eta (0, m_\pi) \ \right]
\ \ \ \ .
\label{eq:practfullamp}
\end{eqnarray}
This expression provides the formally leading energy and 
momentum-transfer dependence of the $\pi d$ scattering amplitude.
The formal EFT expansion parameters are $\gamma/m_\pi$ and $p_\pi/m_\pi$, 
where $p_\pi$ is the pion momentum.
Since $\gamma/m_\pi\sim 1/3$, we expect important higher order 
contributions, arising from the 
momentum dependence 
of the $\pi N$ scattering amplitude (and its effective range expansion) and
also from higher-order $\pi d$ local operators  
involving two or more derivatives, which
are suppressed by additional factors of $\gamma/m_\pi$.

In order to determine
$\tilde\eta (T_\pi, m_\pi) - \tilde\eta (0, m_\pi)$
we integrate the square of the LO amplitude to obtain the cross section for
$\pi d$ scattering, and then enforce unitarity order by 
order in the $c_i$'s.
It is convenient to define the function, $S(x)$, where
\begin{eqnarray}
S(x) & = & 
{2\over x^2}\left[\ 
{1\over 4\gamma^2 r_3^2}
\left(\ 4 \left(\tan^{-1}x\right)^2\ 
\left(\ \log x - i {\pi\over 2}\ \right)
\right.\right.\nonumber\\ & & \left.\left.
\qquad\qquad
\ -\ i 4 \tan^{-1}x \left(\ Li_2(e^{i 2 \tan^{-1}x}) - 
Li_2(-e^{i 2 \tan^{-1}x})\ \right)
\right.\right.\nonumber\\ & & \left.\left.
\qquad\qquad
\ +\ 2 \left(\ Li_3(e^{i 2 \tan^{-1}x}) - 
Li_3(-e^{i 2 \tan^{-1}x})\ \right)
\ -\ {7\over 2}\zeta (3)
\ \right)
\right.\nonumber\\ & & \left.
\qquad
-{2\over\gamma r_3} \left(\ x \tan^{-1}x - {1\over 2}\log\left[ 1 + x^2\right]
\ \right)
\ +\ {x^2\over 2}
\ \right]
\ \ \ ,
\label{eq:Sfun}
\end{eqnarray}
which, by inspection, is real valued, and at $x=0$ has the value 
$S(0)=\left({1-\gamma r_3\over \gamma r_3}\right)^2$.
The functions $Li_n (z)$ are the polylogarithmic functions of order $n$.
To the order we are working, unitarity implies

\begin{eqnarray}
\tilde\eta (T_\pi, m_\pi) - \tilde\eta (0, m_\pi)
& = & 
i 8\pi \left[\ 
{1-\gamma r_3\over \gamma}
(a^+)^2 \sqrt{2 m_\pi T_\pi}
\left(\  1 - 2\left({\gamma r_3\over 1-\gamma r_3}\right)^2
S\left({\sqrt{2 m_\pi T_\pi}\over 2\gamma}\right)
\ \right)
\right.\nonumber\\ & & \left.
\qquad
\ -\ 2 \left[ (a^+)^2 - 2 (a^-)^2\ \right]\tan^{-1}\left({\sqrt{2 m_\pi
      T_\pi}\over 2\gamma}\right)
\ \right]
\ \ \ .
\label{eq:etafix}
\end{eqnarray}
In this expression we have not included contributions of order $\beta$ or
higher as they are higher order in the expansion.
Evaluating eq.~(\ref{eq:practfullamp}) numerically and using 
eqs.~(\ref{eq:Sfun}) and (\ref{eq:etafix}) we find that
the amplitude at threshold is 
${\cal A}_{\pi d} (0, 0)~=~-0.46~{\rm fm}$.  While at $T_\pi = 20~{\rm MeV}$
we find
\begin{eqnarray}
{\cal A}_{\pi d} (20, 0) & = &  \left(\ -0.30 - i 0.00013\ \right)~{\rm fm}
\ \ ,\ \ 
{\cal A}_{\pi d} (20, 150)
\ =\ \left(\ -0.27 - i 0.038\ \right)~{\rm fm}
\ \ \ .
\label{eq:nums}
\end{eqnarray}
Clearly the imaginary part of the amplitude is subleading in the EFT expansion.

\section{Conclusion}

In this work we have shown how to include pions into the pionless effective
field theory.
There are kinematic regimes in processes involving more than one nucleon with
pions in the external states, where there are infrared enhancements resulting
from pions being near their mass-shell in intermediate states.
We have constructed an effective field theory to describe these processes and
have analyzed $\pi d\rightarrow \pi d$. 
The scattering amplitude for this process 
is dominated by the leading non-analytic 
contribution of the form $\sim\log\left(\gamma/m_\pi\right)$, which
we can simply compute in this new effective field theory.

\acknowledgements

We thank Jerry Miller, Iain Stewart and Bira van Kolck for valuable conversations. 
This work is supported in part by the U.S. Dept. of Energy under Grant No.~DE-FG03-97ER4014.

\end{document}